\documentclass{PoS}

\usepackage{amsmath}
\usepackage{graphicx}
\usepackage{subfigure}
\usepackage{bm}

\title{Fluctuation of the initial condition from Glauber models\thanks{Supported by 
Polish Ministry of Science and Higher Education under
grants N202~034~32/0918 and 2~P03B~02828.}}

\ShortTitle{Fluctuations from Glauber models}

\author{\speaker{Wojciech Broniowski}\\
 Institute of Physics, \'Swi\c{e}tokrzyska Academy, PL-25406~Kielce, Poland, and \\
 The H. Niewodnicza\'nski Institute of Nuclear Physics, PL-31342 Krak\'ow, Poland\\
        E-mail: \email{Wojciech.Broniowski@ifj.edu.pl}}

\author{Piotr Bo\.zek\\
   The H. Niewodnicza\'nski Institute of Nuclear Physics, PL-31342 Krak\'ow, Poland, and\\
   Institute of Physics, Rzesz\'ow University, PL-35959 Rzesz\'ow, Poland\\
        E-mail: \email{Piotr.Bozek@ifj.edu.pl}}
        
\author{Maciej Rybczy\'nski\\
 Institute of Physics, \'Swi\c{e}tokrzyska Academy, PL-25406~Kielce, Poland\\
       E-mail: \email{Maciej.Rybczynski@pu.kielce.pl}}      

\abstract{
We analyze measures of the azimuthal asymmetry, in particular 
the {\em participant} harmonic moments, $\varepsilon^\ast$, in a variety of Glauber-like models
for the early stage of collisions at RHIC.
Quantitative comparisons indicate substantial 
model dependence for $\varepsilon^\ast$, reflecting different effective number of sources, 
while the dependence of the scaled 
standard deviation $\sigma(\varepsilon^\ast)/\varepsilon^\ast$ 
on the particular Glauber model is weak. 
For all the considered models the values of  $\sigma(\varepsilon^\ast)/\varepsilon^\ast$ 
range from $\sim$~0.5 for the central collisions to $\sim$~0.3-0.4 for peripheral collisions.
These values, dominated by statistics, change only by 10-15\% from model to model. 
For central collisions and in the absence of correlations between the location of sources
we obtain through the use of the central limit theorem the simple analytic formula 
$\sigma(\varepsilon^\ast)/\varepsilon^\ast(b=0) \simeq \sqrt{4/\pi-1} \simeq 0.52$, independent 
on the collision energy, mass number, or the number of sources.     
We investigate the shape-fluctuation effects for jet quenching and find they are
important only for very central events. 
Finally, we list some remarks and predictions from smooth hydrodynamics 
on higher flow coefficients and their fluctuations, in particular  $\sigma(v_4)/v_4=2\sigma(v_2)/v_2$.}

\FullConference{Critical Point and Onset of Deconfinement
          4th International Workshop\\ July 9-13 2007\\ GSI Darmstadt,Germany}

\begin{document}

\section{Introduction\label{sec:intro}}

This talk is based on Ref.~\cite{BBR}, where more technical details may be found. 

Event-by-event hydrodynamic studies \cite{Aguiar:2000hw,Aguiar:2001ac} 
of relativistic heavy-ion collisions have revealed that {\em fluctuations of the initial shape}
of the system formed in the early stage of the reaction lead to quantitatively relevant 
effects for signatures of the azimuthal asymmetry \cite{Miller:2003kd,Bhalerao:2005mm,Andrade:2006yh,Voloshin:2006gz,volo2}. 
These effects are also important for experimental measurements of the elliptic flow \cite{Alver:2006pn,Alver:2006wh,Sorensen:2006nw,Alver:2007rm}.
In this talk we report our investigation of this phenomenon 
in the framework of various Glauber-like approaches describing the deposition of energy in the 
system in the early stages of the collision. Our study focuses on both 
understanding of the statistical nature of the results, as well as on comparisons of various models.  

\begin{figure}[t]
\begin{center}
\subfigure{\includegraphics[width=.4\textwidth]{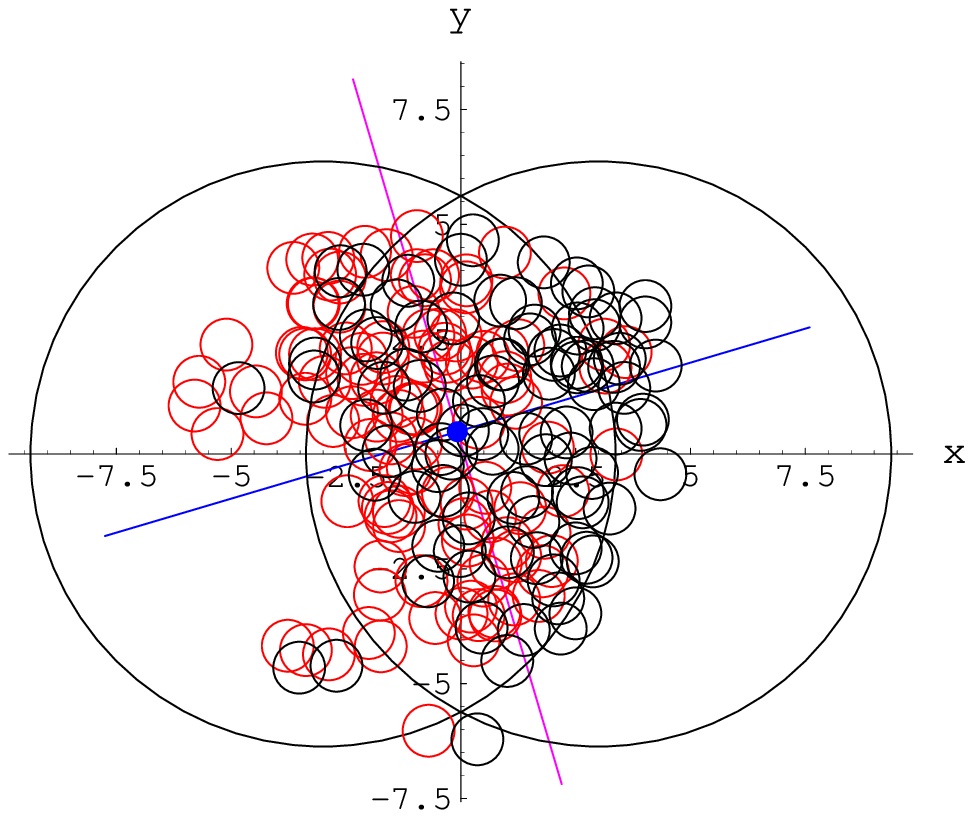}}
\subfigure{\includegraphics[width=.4\textwidth]{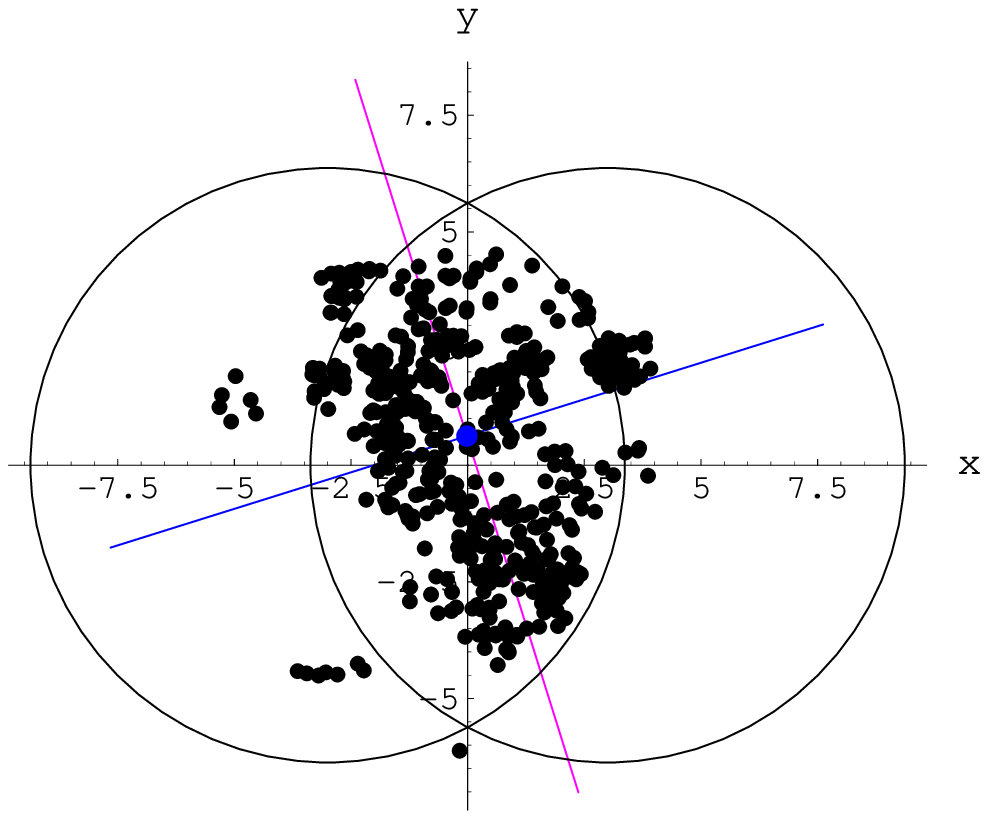}}
\end{center}
\vspace{-5mm}
\caption{A typical gold-gold collision in the $x-y$ plane at $b=6$~fm. 
Left: wounded 
nucleons. Red and black circles indicate 
nucleons from the two colliding nuclei. Right: the centers of mass of pairs of nucleons undergoing binary collisions. The straight lines indicate 
the twisted and shifted principal axes of the quadrupole moment, while the blue dots show the center of mass of the system.\label{fig:snap}}
\end{figure}

Figure~\ref{fig:snap} illustrates the two effects of the shape fluctuations due to the finite number of 
sources: the shift of the center-of-mass and the rotation of the the quadrupole principal axes.
Statistical analyses may be performed in the reference frame 
fixed by the reaction plane (we call it {\em fixed-axes}, a.k.a. standard), or for each event in the frame defined by the twisted and shifted principal axes
(we call it the {\em variable-axes} frame, a.k.a. participant). 
In the fixed-axes frame the two-dimensional probability distribution of sources can be Fourier-expanded as
\begin{eqnarray}
f(\rho,\phi)=f_0(\rho) + 2 f_2(\rho) \cos(2\phi)+ 2 f_4(\rho) \cos(4\phi)+ \dots, \label{f}
\end{eqnarray}
where the transverse radius $\rho=\sqrt{x^2+y^2}$ is measured from the center of the geometric intersection of the two nuclei. One also introduces \vspace{-3mm}
\begin{eqnarray}
\varepsilon_{l}=\frac{\int 2\pi \rho f_l(\rho) \rho^2 d\rho }{\int 2\pi \rho f_0(\rho) \rho^2 d\rho }.
\label{epsilon}
\end{eqnarray}
On the other hand, in the variable-axes frame we have the distribution
\begin{eqnarray}
f^\ast(\rho,\phi)=f_0(\rho) + 2 f_2^\ast(\rho) \cos(2\phi-2\phi^\ast)+ 2 f_4^\ast(\rho) \cos(4\phi-4\phi^\ast)+ \dots, \label{fv}
\end{eqnarray}
where $\phi^\ast$ denotes the rotation angle of the principal axes in each event. Correspondingly,
\begin{eqnarray}
\varepsilon_{l}^\ast=\frac{\int 2\pi \rho f_l^\ast(\rho) \rho^2 d\rho }{\int 2\pi \rho f_0(\rho) \rho^2 d\rho } 
\label{epsilonv}
\end{eqnarray}
($f_0=f_0^\ast$). 
The quadrupole parameters are denoted without the subscript as $\varepsilon=\varepsilon_2$ and $\varepsilon^\ast=\varepsilon_2^\ast$.

\section{The toy problem\label{sec:toy}}

Consider the one-dimensional problem where  uncorrelated particles are randomly generated from
a distribution in the azimuthal angle $\phi$ containing the monopole and quadrupole moments,
\begin{eqnarray}
f(\phi) = 1+2\epsilon \cos(2 \phi), \;\;\;\;\; \epsilon \in [-\frac{1}{2},\frac{1}{2}]. \label{toydist}
\end{eqnarray} 
Obviously, the distribution has only two non-zero fixed-axes moments, 
\begin{eqnarray}
f_0=\frac{1}{2\pi} \int_0^{2\pi} d\phi f(\phi)=1, \;\;\;\;\;
f_2=\frac{1}{2\pi} \int_0^{2\pi} d\phi \cos(2 \phi) f(\phi)=\epsilon.
\end{eqnarray}
We generate $n$ particles according to the distribution (\ref{toydist}) in each event, and 
subsequently carry out the averaging over the events, denoted as $\langle \langle . \rangle \rangle$.
For instance, $f_2$ is estimated as 
\begin{eqnarray}
f_2 \simeq  \langle \langle \frac{1}{n}\sum_{k=1}^n \cos(2 \phi_k) \rangle \rangle,
\end{eqnarray}
where $k$ labels the particles the event. The equality becomes strict as the number of events approaches 
infinity, which is assumed implicitly.
In the variable-axes case we rotate 
the particles by the angle $\phi^\ast$ in each event. Thus \vspace{-4mm}
\begin{eqnarray}
f_2^\ast \equiv \varepsilon^\ast = \langle \langle \frac{1}{n}\sum_{k=1}^n \cos[2 (\phi_k-\phi^\ast)] \rangle \rangle. \label{toyf2r}
\end{eqnarray}
The rotation angle $\phi^\ast$ depends itself on the distribution of particles in the given event. By definition, it is chosen in such a way that 
the quantity $\frac{1}{n}\sum_{k=1}^n \cos[2 (\phi_k-\phi^\ast)]$ assumes maximum, which gives the conditions \vspace{-4mm}
\begin{eqnarray}
&& \cos(2\phi^\ast)=Y_2/\sqrt{Y_2^2+X_2^2}, \;\;\;\;\;\; \sin(2\phi^\ast)=X_2/\sqrt{Y_2^2+X_2^2}, \label{toysums} \\
&&Y_2=\frac{1}{n}\sum_{k=1}^n \cos(2 \phi_k), \;\;\;\;\;\; X_2=\frac{1}{n}\sum_{k=1}^n \sin(2 \phi_k). \nonumber
\end{eqnarray}
Using the above formulas in Eq.~(\ref{toyf2r}) yields 
\begin{eqnarray}
f_2^\ast &=&  \langle \langle \sqrt{Y_2^2+X_2^2} \rangle \rangle \label{toys} 
= \langle \langle \sqrt{\left ( \frac{1}{n}\sum_{k=1}^n \cos(2 \phi_k) \right )^2+ 
\left ( \frac{1}{n}\sum_{k=1}^n \sin(2 \phi_k) \right )^2} \rangle \rangle . 
\end{eqnarray}
We see that the variable-axes moment corresponds to an average of the square root of sums (\ref{toysums}), thus is 
a highly ``non-local'' object, involving upon expansion infinitely many fixed-axes moments.

\begin{figure}[tb]
\begin{center}
\subfigure{\includegraphics[width=.42\textwidth]{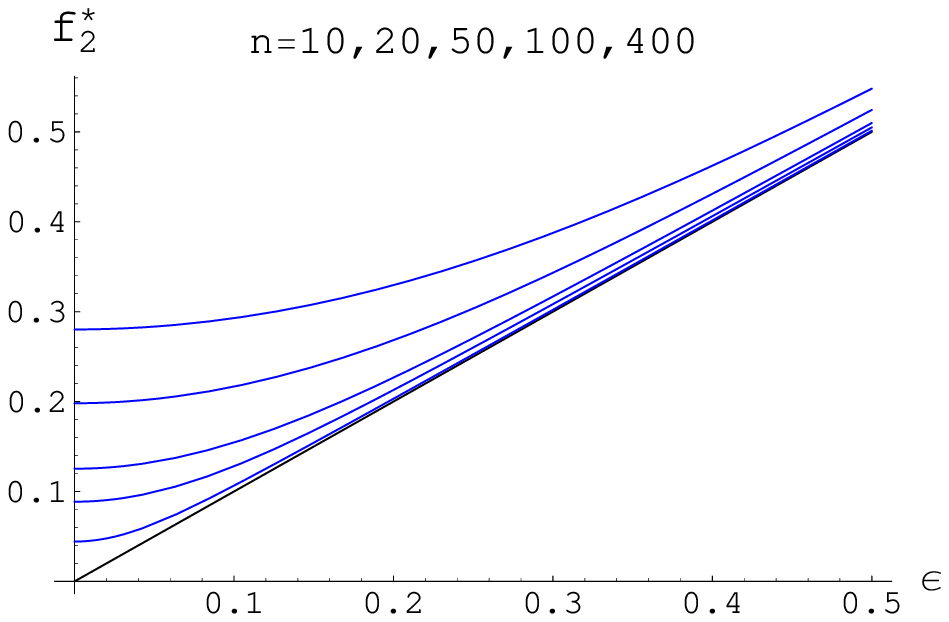}}
\subfigure{\includegraphics[width=.42\textwidth]{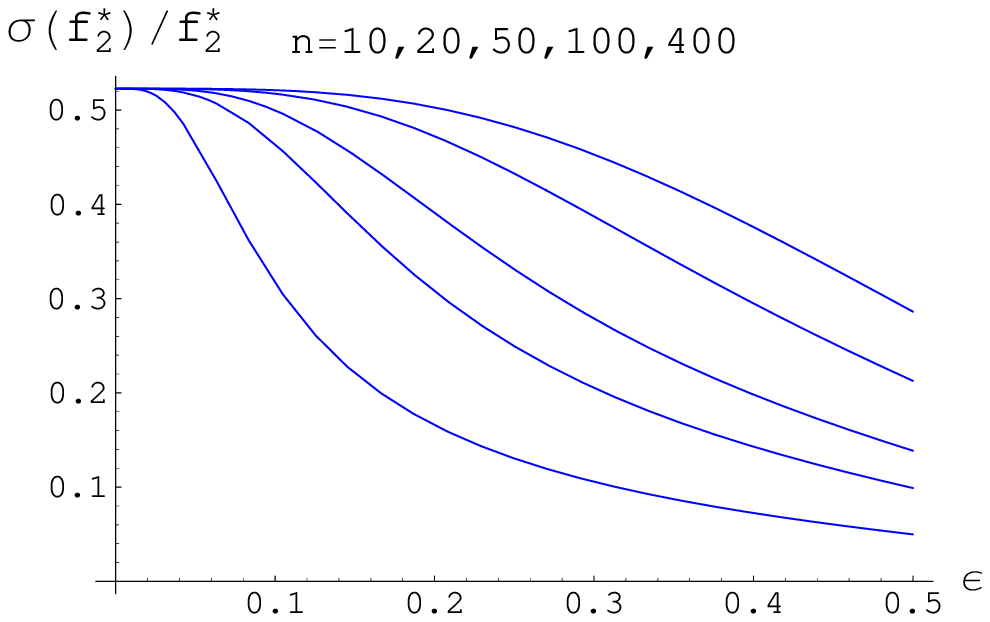}} 
\end{center}
\vspace{-4mm}
\caption{Toy model. Left: Dependence of the variable-axes moment $f_2^\ast$ on the fixed-axes 
quadrupole moment $\epsilon$ 
for several values of the number of particles $n$. As $n$ increases, we pass from top to bottom with 
the presented curves. The straight line is the $n \to \infty$ limit, {\em i.e.} $f_2^\ast=\epsilon$. 
Right: the same for the scaled standard deviation.
\label{fig:dexp}}
\end{figure}

For sufficiently large  $n$ one may evaluate Eq.~(\ref{toys}) with the help of the {\em central 
limit theorem}. Consider the variables $c_k = \cos (2 \phi_k)$ and \mbox{$s_k = \sin (2 \phi_k)$}. Their averages and variances are 
\begin{eqnarray}
\bar c =\epsilon, \;\;\sigma^2_c =\frac{1}{2}-\epsilon^2,  \;\;
\bar s = 0, \;\; \sigma^2_s = \frac{1}{2}. 
\end{eqnarray}  
Importantly, there is no correlation between $Y_2$ and $X_2$, as $\frac{1}{2\pi}\int_0^{2\phi} d\phi \cos(2\phi) \sin(2\phi) f(\phi)=0$.
Thus, according to the central limit theorem, the distribution of $Y_2$ and $X_2$ is Gaussian.
Introducing 
\begin{eqnarray}
Y_2=q \cos \alpha, \;\;\;\; X_2=q \sin \alpha, \;\;\;\; q^2=Y_2^2+X_2^2, \;\;\;\; 
\delta=\frac{1}{2\sigma_c^2}-\frac{1}{2\sigma_s^2} = \frac{1}{1-2\epsilon^2} -1, \label{toydefs}
\end{eqnarray}
we may write this distribution in the form 
\begin{eqnarray}
f(X_2,Y_2)= f(q,\alpha)=\frac{n}{\pi \sqrt{1-2\epsilon^2}} \exp{\left [ -n \left ( \frac{q^2+\epsilon^2-2 q \epsilon \cos \alpha}{1-2\epsilon^2} \right ) + n \delta q^2 \sin^2\alpha \right ]}. 
\end{eqnarray}
We need  below the integral of this distribution over $\alpha$, which can be expanded as \cite{Poskanzer:1998yz,Sorensen:2006nw}
\begin{eqnarray}
\!\!\!\!\!\! \int_0^{2\pi} \!\!\!\! d\alpha f(q,\alpha)\!=\!\frac{2n}{\sqrt{\pi} \sqrt{1\!-\!2\epsilon^2}} 
\exp{\left [ -n \left ( \frac{q^2+\epsilon^2}{1-2\epsilon^2} \right ) \right ]} 
\sum_{j=0}^\infty \left ( 2 q \epsilon \right )^j \frac{\Gamma(j+\frac{1}{2})}{j!}
I_j \left ( \frac{2 n \epsilon q}{1-2\epsilon^2}\right ), \label{mean}
\end{eqnarray}
where $I_j(x)$ are the modified Bessel functions.
We may now express Eq.~(\ref{toyf2r}) as the series involving the confluent hypergeometric function,
\begin{eqnarray}
\!\!\!\!\!\!\!\!\!\!\!\!\!\! f_2^\ast\!=\!\int \!\!q\,dq \,d\alpha \, q f(q,\alpha) \!=\! \frac{1-2\epsilon^2}{\sqrt{n \pi}} 
 \sum_{j=0}^\infty \!\left ( 2\epsilon^2 \right )^j \frac{\Gamma(j+\frac{1}{2})\Gamma(j+\frac{3}{2})}{j!^2}
{}_1F_1 \!\left (-\frac{1}{2},j\!+\!1;-\frac{n \epsilon^2}{1\!-\!2\epsilon^2} \right ), \label{1F1}
\end{eqnarray}
which converges fast and can be used for practical calculations in a truncated form.
At $\epsilon=0$ (azimuthally symmetric distribution) we have the very simple result
\begin{eqnarray}
f_2^\ast(\epsilon=0)&=&\frac{\sqrt{\pi}}{2 \sqrt{n}},
\end{eqnarray}
which shows the expected $1/\sqrt{n}$ behavior for a statistical fluctuation. 
The numerical results obtained with the series (\ref{1F1}) are presented in Fig.~\ref{fig:dexp}, left side.  
We note that the effect of the departure of $f_2^\ast$ from $\epsilon$ is strongest at 
low $\epsilon$ and low $n$.

The evaluation of the second moment in the $q$ variable yields
\begin{eqnarray}
\int q\, dq \,d\alpha \, q^2 f(q,\alpha) = \frac{1+(n-1) \epsilon^2}{n}.
\label{secmom} 
\end{eqnarray}
From Eqs.~(\ref{1F1},\ref{secmom}) we can now obtain the variance of the distribution of the variable-axes moment.
Again, a simple formula follows for the case $\epsilon=0$, where 
${\rm var}(f_2^\ast)=(1-\frac{\pi}{4})/{n}.$
The scaled variance and scaled standard deviation are
\begin{eqnarray}
\frac{{\rm var}(f_2^\ast)}{f_2^\ast}=\frac{\frac{2}{\sqrt{\pi}}-\frac{\sqrt{\pi}}{2}}{\sqrt{n}}, 
\;\;\;\;\;\;\;\;\;\;\;
\frac{\sigma(f_2^\ast)}{f_2^\ast}=\sqrt{\frac{4}{\pi}-1} \simeq 0.523, \;\;\;\;\;\; (\varepsilon=0). \label{estimates}
\end{eqnarray}
Note that in this case there is no dependence of the scaled standard deviation on $n$.
The case of general $\varepsilon$ obtained numerically 
for various values of $n$ is shown in Fig.~\ref{fig:dexp}, right side. According to Eq.~(\ref{estimates}), 
all curves approach the limit $\sqrt{\frac{4}{\pi}-1}$ as $\varepsilon \to 0$. At the other end, in 
the limit of $n \varepsilon^2 \to \infty$ we have 
the expansions $f_2^\ast = \varepsilon+1/(4 \varepsilon n)+\dots$ and 
$\sigma(f_2^\ast)/f_2^\ast = [1/(2\varepsilon)-\varepsilon]/n+\dots$.

\section{The general case\label{sec:moments}}

In the general case the analysis can be carried out in full analogy to the toy model \cite{BBR}.
For simplicity, in our analytic study we neglect correlations between locations of sources.
If such correlations are strong, 
their analytic inclusion is difficult and one has to resort to numerical simulations such as those presented below. 
Compared to the  toy model, the full two-dimensional case involves the fixed-axes moments
\mbox{$I_{k,l}= \int_0^\infty 2\pi \rho d\rho f_l(\rho) \rho^k/n$},
where $n$ is the number of sources. We have chosen the normalization \mbox{$\int_0^\infty 2\pi \rho d\rho f_0(\rho)=n$}. 
Generally, in analogy to Eq.~(\ref{1F1}) \vspace{-3mm}
\begin{eqnarray}
\varepsilon^\ast = \frac{\sqrt{2} \sigma_{Y_2}^2}{I_{k,0} \sqrt{\pi } \sigma_{X_2}} \sum _{m=0}^{\infty } (2 \delta \sigma_{Y_2}^2)^m  
 \label{epsv} \frac{
\Gamma \left(m+\frac{1}{2}\right) \Gamma \left(m+\frac{3}{2}\right) \,
   _1F_1\left(-\frac{1}{2};m+1;-\frac{\bar Y_2^2}{2 \sigma_{Y_2}^2}\right)}{ m!^2}, 
\end{eqnarray}
where \vspace{-3mm}
\begin{eqnarray}
\bar Y_2=I_{k,2}, \;\; \sigma^2_{Y_2}= \frac{1}{2n}(I_{2k,0}-2 I_{k,2}^2+ I_{2k,4}), \;\;
\sigma^2_{X_2}= \frac{1}{2n}(I_{2k,0}-I_{2k,4}), \;\; 
\delta = \frac{1}{2 \sigma_{Y_2}^2}-\frac{1}{2 \sigma_{X_2}^2}.
\end{eqnarray}
For the special case of central collisions we have
the very simple results
\begin{eqnarray}
\varepsilon^\ast = \frac{\sqrt{\pi I_{2k,0}} }{2I_{k,0} \sqrt{n}},\;\;\;\;\; 
\frac{\sigma ( \varepsilon^\ast)}{\varepsilon^\ast}=\sqrt{\frac{4}{\pi}-1}\simeq 0.523, \;\;\;\;\;\; (b=0). \label{ratioeps}
\end{eqnarray}
Since correlations between the location of sources effectively reduce the number of sources $n$, 
they lead to an increase of $\varepsilon^\ast$, but 
keep its $n$-independent scaled variance practically constant, as shown by the simulations of the next Section.  
Ref.~\cite{BBR} contains more discussion.

\section{Numerical simulations in various Glauber models \label{sec:models}}

\begin{figure}
\subfigure{\includegraphics[width=.5\textwidth]{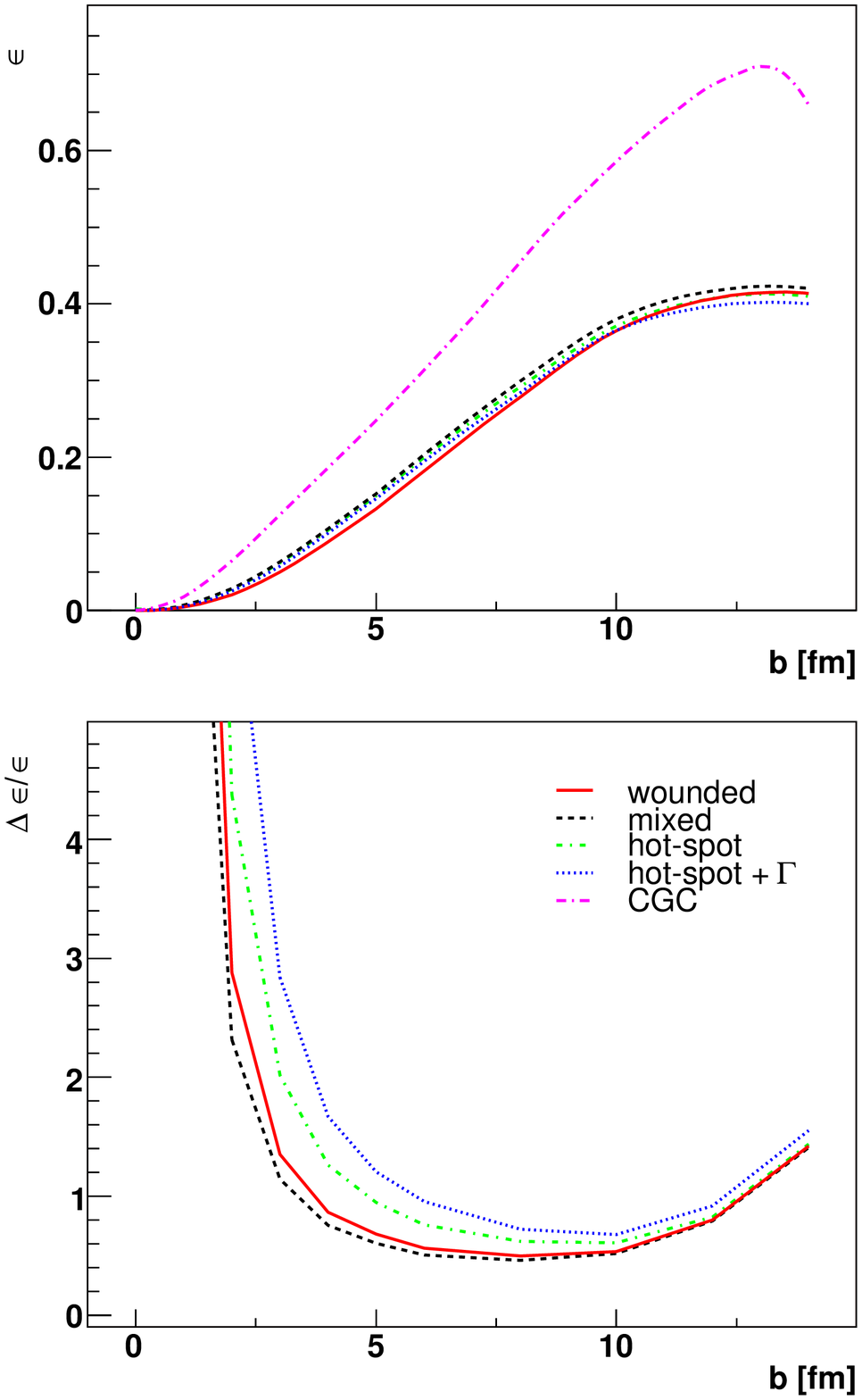}}
\subfigure{\includegraphics[width=.5\textwidth]{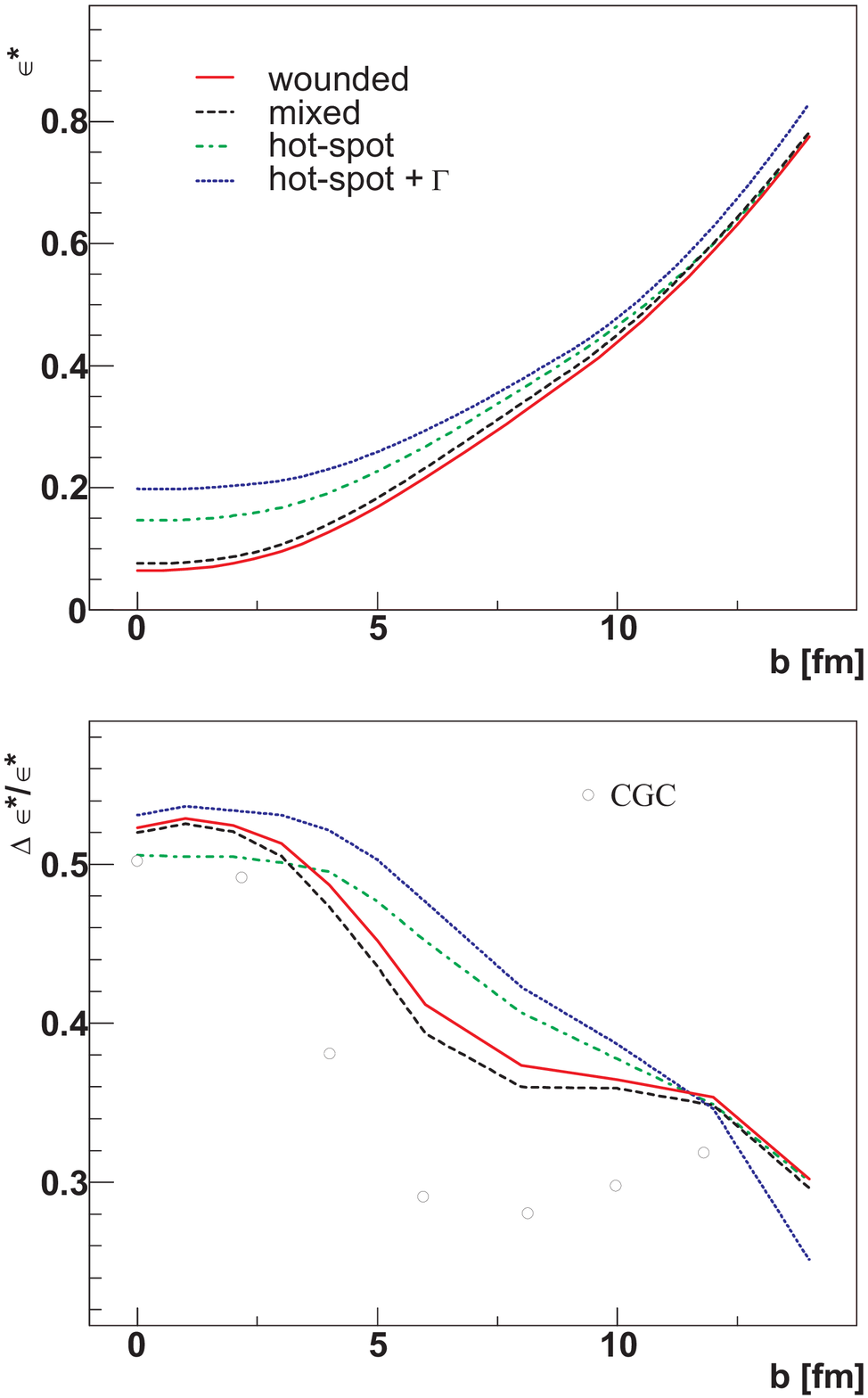}}
\caption{Left: The moment $\varepsilon$ and its scaled standard deviation 
for the analyzed models
plotted as functions of the impact parameter. Gold-gold collisions. Right: the same for $\varepsilon^\ast$. 
The results for the color-glass condensate come from Ref.~\cite{Hirano:2005xf} (dot-dashed line in top left figure) and Ref.~\cite{DNara} 
(circles in lower right figure). \label{fig:epsilons}}
\end{figure}

We have studied a few variants of Glauber-like models.
In the standard 
wounded nucleon model  \cite{Bialas:1976ed} the weight $w=1/2$ is attributed to the point 
in the transverse plane at the position of the wounded nucleon. The wounding cross section is $42$~mb.
For binary collisions the weight $w=1$ is attributed to each collision point.
We remark that only relative magnitude of weights is important in 
studies of fluctuations.
A successful description of the RHIC multiplicities has been achieved with a {\em mixed} model, 
amending wounded nucleons 
with some admixture of binary collisions \cite{Back:2001xy,Back:2004dy}. 
Then the wounded nucleon obtains the weight $w=(1-\alpha)/2$, and the binary collision 
the weight $w=\alpha$. The total weight averaged over events 
is then $(1-\alpha)N_{\rm w}/2+\alpha N_{\rm bin}$. The fits to particle 
multiplicities of Ref.~\cite{Back:2004dy} give $\alpha = 0.145$ at $\sqrt{s_{NN}}=200$~GeV.
We also consider a model with {\em hot spots} in the spirit of Ref.~\cite{Gyulassy:1996br}, assuming that the 
cross section for a semi-hard binary collisions producing a hot-spot 
is tiny, $\sigma_{\rm hot-spot} = 0.5$~mb, however when such a rare collision occurs it produces 
on the average a very large amount of the transverse energy equal to $\alpha\sigma_{\rm w}/\sigma_{\rm hot-spot}$. 
Each source from the previously described models
deposits the transverse energy with a certain probability distribution. To incorporate this effect, we
superimpose the $\Gamma$ distribution, multiplying the weights of the considered model with the 
randomly distributed number from the gamma distribution 
\mbox{$g(w,\kappa)={w^{\kappa-1}\kappa^\kappa \exp(-\kappa w)}/{\Gamma(\kappa)}$.} 
Here we do this superposition on the hot-spot model, labeled {\em hot-spot+$\Gamma$}.
Thus, we take the weights $(1-\alpha)g(w,\kappa)/2$ for the wounded nucleons and  $\alpha g(w,\kappa)\sigma_{\rm w}/\sigma_{\rm hot-spot}$ 
for the binary collisions. We set $\kappa=0.5$, which gives ${\rm var}(w)=5$.
The four considered models (wounded nucleon, mixed, hot-spot, and hot-spot+$\Gamma$) 
differ substantially by the number of sources and the amount of the built-in fluctuations. 
    
We observe that in all four models $\varepsilon$ is practically independent of the model (top left panel of 
Fig.~\ref{fig:epsilons}). 
On the other hand, the scaled standard deviation (lower left
panel of Fig.~\ref{fig:epsilons}) displays a strong dependence on the model at low values of $b$,
with the hot-spot+$\Gamma$ model yielding about twice as much as the mixed model. 
We also notice a very strong dependence on $b$. At $b=0$ the curves diverge due to dividing by the vanishing 
value of $\varepsilon$.
The fluctuations are larger in models effectively having the lower number of sources, which is obvious 
from the statistical point of view.

As already noted in Refs.~\cite{Hirano:2005xf,Drescher:2006ca}, 
the value of $\varepsilon$ obtained with the color glass condensate (CGC)  
is substantially higher than in all Glauber-like models reported here (upper curve in the left top panel of Fig.~\ref{fig:epsilons}).

The quadrupole moment $\varepsilon^\ast$ and its scaled standard deviation are show on the right side of 
Fig.~\ref{fig:epsilons} 
We observe a strong model dependence of $\varepsilon^\ast$ at low values of $b$, with models having effectively lower number 
of sources yielding higher values. At $b=0$ the hot-spot+$\Gamma$ model yields three times more than the 
wounded nucleon model. For all models the scaled standard deviation is close to the value
0.5 for central collisions (in agreement with the results (\ref{ratioeps})) and drops to about 0.3 at $b=14$~fm. 
At intermediate values of $b$ the relative difference in $\sigma(\varepsilon^\ast)/\varepsilon^\ast$ between various
considered models is at the level of 10-15\%, which is not a very strong effect. 
The CGC result of Ref.~\cite{DNara} is lower than in the Glauber 
models (circles in the right bottom panel of Fig.~\ref{fig:epsilons}).

The harmonic profiles $f_l(\rho)$ and $f^\ast_l(\rho)$ are displayed in Ref.~\cite{BBR}

\section{Jet quenching\label{sec:jets}}

\begin{figure}[b]
\begin{center}
\subfigure{\includegraphics[width=.27\textwidth]{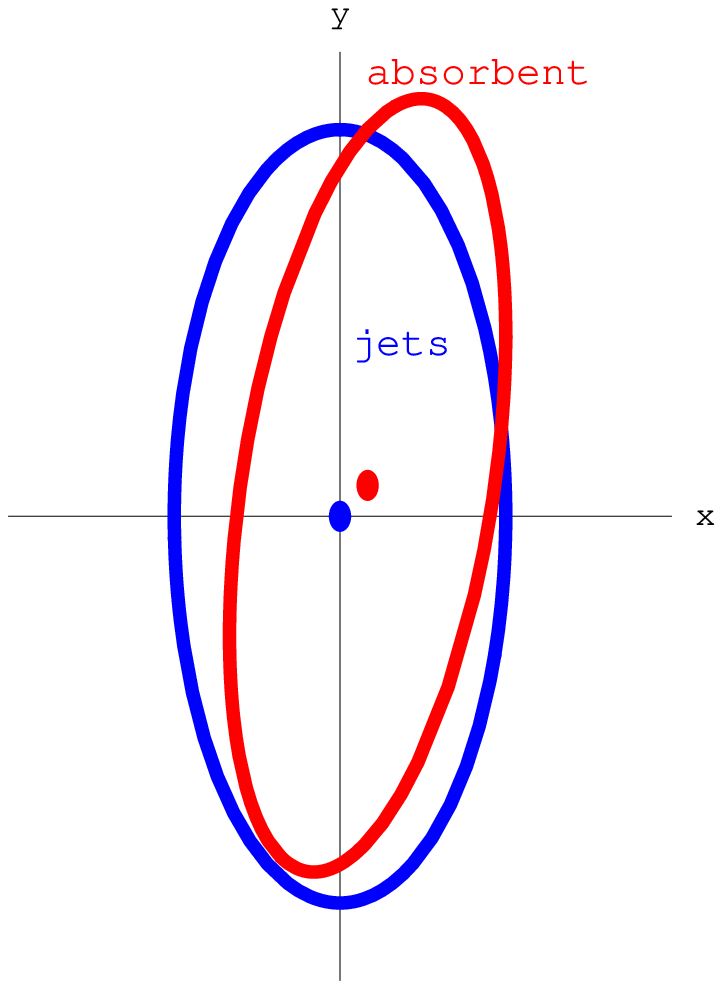}}
\subfigure{\includegraphics[width=.48\textwidth]{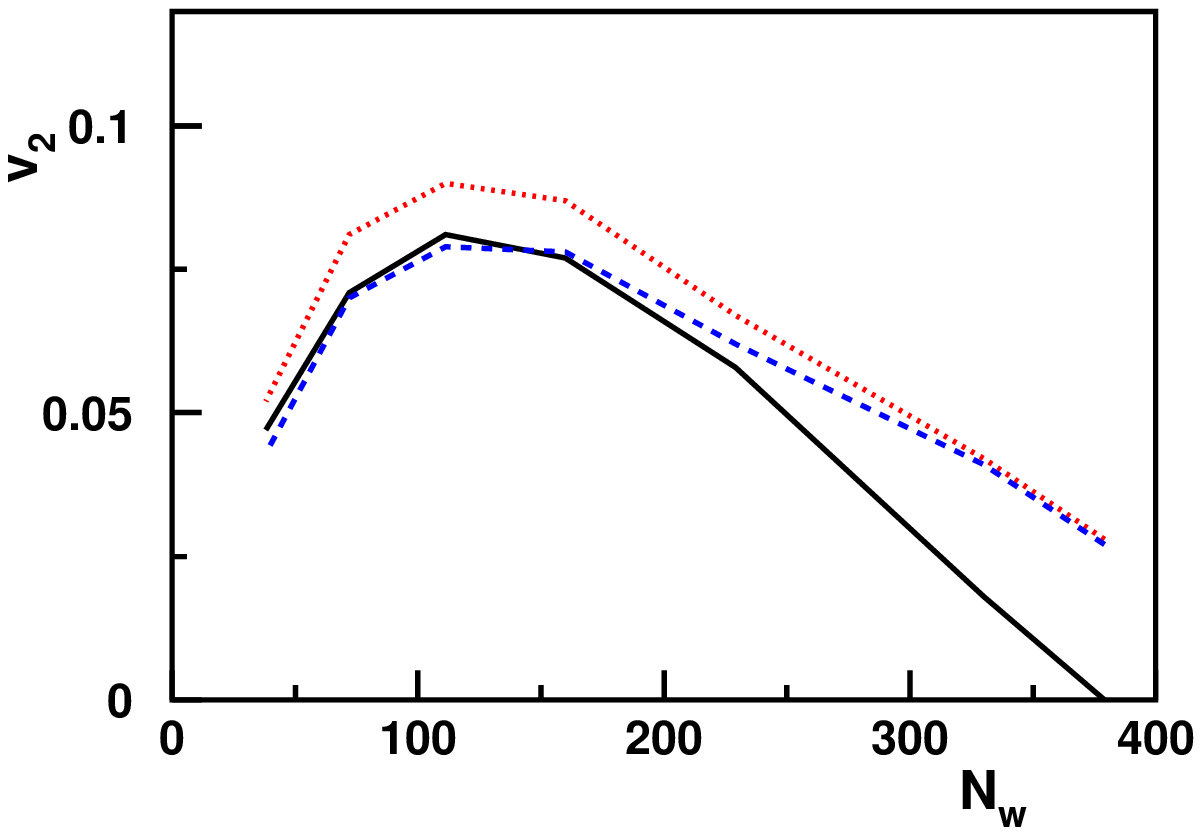}}
\end{center}
\vspace{-6mm}
\caption{Left: the variable axes geometry of the absorbent vs. the fixed-axes geometry of the jet production point. 
Right: $v_2$ at high $p_T$ as a function of the number of wounded
nucleons, obtained with the variable-axes density $f^\ast(\rho)$ 
for the hot-spot scenario (dashed line), and  
with the fixed-axes density of the wounded nucleons $f(\rho)$  (solid line). The dotted line represents the 
result for the variable-axes density but without the shift and rotation of the opaque medium.  
\label{fig:v2}}
\end{figure}

We have used the model of Refs.~\cite{Drees:2003zh,Horowitz:2005ja} of the jet energy loss in order to explore
the role of the event-by-event rotated absorbing medium.
In order to take into account the variable-axes geometry, we use
$f^\ast(\rho,\phi)$ as the density of the scattering centers for the
propagating parton. The rare jet production event is distributed according to the fixed-axes profile, see 
Fig.~(\ref{fig:v2}), left side. The resulting increase of the eccentricity of the absorbent 
is expected to increase the asymmetry of the jet absorption. 
A very similar effect has been discussed for profiles calculated 
in the CGC model \cite{Drescher:2006pi}, where an increase in $v_2$ by about $10-15\%$ has been found.
The absorbing medium formed in each event is rotated and also shifted.
The elliptic flow (see Fig.~\ref{fig:v2}, right side) at centralities larger than $20\%$
calculated with the wounded-nucleon model in the fixed-axes frame (solid line), which serves as a reference,
comes out similar to the 
result of the hot-spot model in the variable-axes frame (dashed line).
Only if the shift and rotation of the opaque medium were neglected (dotted line)
the modification of the shape leads to an increase of the high $p_T$
elliptic flow coefficient $v_2$ by about $10-15\%$. 
The cancellation of the effects of the increased 
eccentricity of the medium and of the shift and rotation happens also for the other considered models
(at larger centralities). 
The rotation of the absorbing medium yields about 2/3, and the shift about 1/3 of the 
total cancellation effect.

\section{Fluctuations of the elliptic flow\label{sec:flow}}

\begin{figure}[b]
\begin{center}
\includegraphics[width=.47\textwidth]{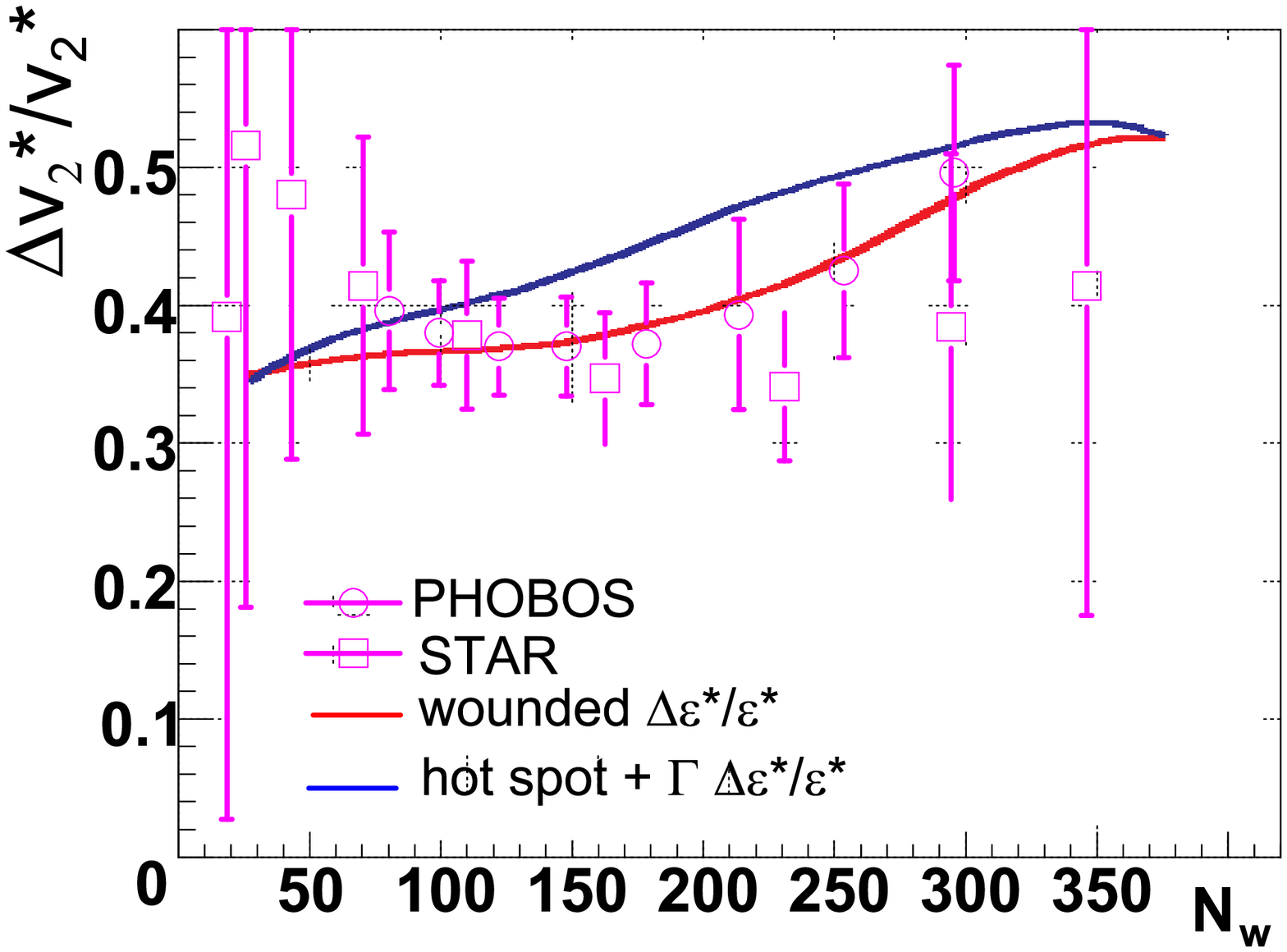}
\end{center}
\vspace{-7mm}
\caption{Fluctuations of $v_2^\ast$. Data from Refs.~\cite{Alver:2006wh,Sorensen:2006nw,Alver:2007rm}.\label{fig:v22}}
\end{figure}

The fluctuations of the elliptic flow, which are an important 
probe of the nature of the early-stage dynamics of the system 
\cite{Mrowczynski:2002bw}, have recently been measured at RHIC
\cite{Alver:2006wh,Sorensen:2006nw,Alver:2007rm}. The experimental 
procedure used in these analyses identifies the elliptic flow coefficient with 
the variable axes $v_2$, here denoted as $v_2^\ast$.
The relevance of studies of fluctuations of the initial shape  comes from the 
well-known fact that 
for small elliptic asymmetry one expects on hydrodynamic grounds the relation \vspace{-3mm}
\begin{eqnarray}
\frac{\sigma( v_2^\ast)}{v_2^\ast}=\frac{\sigma( \varepsilon^\ast)}{\varepsilon^\ast}. \label{hydrofl}
\end{eqnarray}
As argued in Ref.~\cite{Vogel:2007yq}, the result (\ref{hydrofl})
indicates that the mean free path in the matter created in the initial stages of the heavy-ion collisions
is very small, although turbulence does not develop. Comparison of the data to our Glauber calculations
is made in Fig.~\ref{fig:v22}. For central collisions we expect \vspace{-3mm}
\begin{eqnarray}
\frac{\sigma(v_2^\ast)}{v_2^\ast}(b=0)\simeq\frac{\sigma(\varepsilon^\ast)}{\varepsilon^\ast}(b=0)\simeq\sqrt{\frac{4}{\pi}-1} \simeq 0.52, \label{v2res}
\end{eqnarray}
which is compatible to the data, although the error bars are large.

We end this talk with some comments on the derivation of Eq.~(\ref{hydrofl}) as well as on higher-harmonic probes. 
Perturbation theory applied to smooth ({\em i.e.} linearization is sensible) hydrodynamics together with hierarchy of relaxation times for 
subsequent harmonics, {\em e.g.} $\tau_2 \gg  \tau_4$, leads  
to further results \cite{BBR}. In particular, \vspace{-3mm}
\begin{eqnarray}
v_4^\ast \sim \varepsilon^{\ast 2} \sim v_2^{\ast 2}. \label{ev2}
\end{eqnarray}
In Ref.~\cite{Borghini:2005kd} the variable $v_4/v_2^{2}$ has 
been suggested as a sensitive probe of the hydrodynamic 
evolution. The simulations of Refs.~\cite{Kolb:2003zi,Borghini:2005kd} show that with increasing 
time the value of $v_2$ saturates, while $v_4$ quickly 
assumes the value proportional to $v_2^{2}$, supporting the assumption $\tau_2 \gg  \tau_4$ 
used in the above argumentation. 
For the fluctuations one gets immediately from Eq.~(\ref{ev2}) the prediction \vspace{-3mm}
\begin{eqnarray}
\frac{\sigma(v_4^\ast)}{v_4^\ast} = 2\frac{\sigma(v_2^\ast)}{v_2^\ast}. \label{ev2fl}
\end{eqnarray}
Relation (\ref{ev2fl}), if verified experimentally, would support the scenario of smooth hydro 
evolution with the mentioned hierarchy of scales. 
On similar grounds, 
for the azimuthal Hanbury-Brown--Twiss (HBT) correlation radius, $R^{\rm HBT}(\phi)$, one expects \vspace{-3mm}
\begin{eqnarray}
R^{\rm HBT}_4  \sim (R^{\rm HBT}_2)^2, \label{HBT}
\end{eqnarray} 
where $R^{\rm HBT}(\phi)=R^{\rm HBT}_0+2 R^{\rm HBT}_2 \cos(2 \phi)+2R^{\rm HBT}_4 \cos(4 \phi) +\dots$. 
%The proportionality constant in Eq.~(\ref{HBT}) is the same as in Eq.~(\ref{ev2}).

\section{Conclusion\label{sec:concl}}

Here are our main points:

\begin{itemize}

\item We have analyzed four Glauber-like models, with different degree of fluctuation: 
the wounded-nucleon model, the mixed model, the hot-spot model, and the hot-spot model with the 
superimposed $\Gamma$ distribution.

\item We have obtained numerically the fixed-axes and variable-axes harmonic profiles \cite{BBR} and analyzed their moments. 
The variable-axes moments $\varepsilon^\ast$, and 
the fixed-axes scaled standard deviation $\sigma(\varepsilon)/\varepsilon$ are sensitive to the choice of the model, while 
$\sigma(\varepsilon^\ast)/\varepsilon^\ast$ is not, changing at most by 10-15\% from model to model at intermediate values of $b$. 

\item Analytic formulas explain certain 
features of the simulations, in particular, they show that at $b=0$ the multiple-axes scaled variances are 
close to the value 0.5, insensitive of the model used, the collision energy, the mass number of the colliding nuclei, or the 
number of particle sources. The 
behavior of $\sigma( \varepsilon^\ast)/\varepsilon^\ast$ at low $b$ is thus largely {\em governed by the statistics}

\item Fixed-reaction-plane experimental analyses would reveal more information on the system and would allow to 
discriminate the theoretical predictions, as fluctuations of $\varepsilon$ are sensitive to the chosen model. 
 
\item  For the jet $v_2$  we find that the effect of the increased variable-axes eccentricity is largely canceled by the shift of the center 
of mass and the rotation of the principal axes of the absorbing medium. This leads to practically no change 
of the jet emission asymmetry at intermediate and large impact parameters. Only at small $b$ the increase 
of the quadrupole moment takes over the relatively less important shift and rotation.
 
\item On hydrodynamic grounds, the analysis of the variable-axes moments in the coordinate space 
carries over to the collective flow and analysis of $v_2^\ast$. In particular, 
Eq.~(\ref{v2res}) holds for the variable-axes elliptic flow coefficient. 

\item Under assumptions of smoothness, perturbation theory made on top of azimuthally symmetric hydro leads to 
sensitivity of higher-harmonic late-time measures, $v_4^\ast$, $R^{\rm HBT}_4$, {\em etc.}, to the initial {\em quadrupole} deformation 
$\varepsilon^\ast(t_0)$ only. 
Higher harmonics of the initial shape deformation are irrelevant, as they presumably are damped fast. 
A number of relations follows for various measures and their event-by-event fluctuations, {\em e.g.} Eq.~(\ref{ev2fl}).

\item It would be a challenge to measure the $v_4^\ast$ fluctuations and test the smooth hydro assumption 
by verifying relation  (\ref{ev2fl}).

\end{itemize}

One of us (WB) thanks Paul Sorensen, Constantin Loizides, and Wit Busza
for useful discussions concerning the experimental determination of $v_2$ and its fluctuations. 

%\bibliographystyle{elsart-num}
%\bibliography{eps}

\end{document}